# Pseudo-periodicity and 1/*f* noise from the sum of similar intermittent signals

Giovanni Zanella

Dipartimento di Fisica dell'Università di Padova and Istituto Nazionale di Fisica Nucleare, Sezione di Padova, via F. Marzolo, 8, 35131 Padova, Italy

**Abstract**

The usual interpretation of $1/f$ noise is represented by a sum of many independent two-level elementary random signals with a distribution of relaxation times. In this paper it is demonstrated that also the superposition of many similar single-sided two-level signals, with the same relaxation time, produces $1/f$ noise. This is possible tanks to the coincidences among the signals which introduce cross-correlations and tune locally the resulting process in trains of pseudo-periodic pulses. Computer simulations demonstrate the reliability of this model, which permits to insert in an coherent framework other models solving problems still open.

In spite of its ubiquity, a general physical mechanism for $1/f$ noise has not be accepted yet. However, the study of the intermittent signals, describing the random presence of the same object, has provided a powerful means to investigate $1/f$ noise [1]. These signals can be modeled by the so-called random telegraph signals (RTSs), which are characterized by a relaxation time (the inverse of the total rate of transitions back and forth in the process) and by a Lorentzian power spectrum [2]. The usual interpretation of the $1/f$ noise by a sum of several independent Lorentzians, with different relaxation times [1][3], is not convincing at all, due to lack of generality and absence of cross-correlation contributions [4]. On other hand, the generation of $1/f$ noise by a superposition of similar single-sided RTSs has been always excluded, due the result of another Lorentzian and a steady term representing averagely the cross-correlation contributions [5]. Here, we show that summing many similar single-sided RTSs, the cross-correlation contributions cannot be assimilate to a steady level, but they cause trains of pseudo-periodic pulses generating $1/f$ noise. Indeed, the signal fluctuations emerging from the mean level on the resulting signal, being generated by coincidences in up or down level among the RTSs, tune locally the resulting process in pulse-trains, which length depends essentially on the amplitude of the dominant fluctuation.

According to the usual interpretation of $1/f$ noise as due to a superposition of elementary signals, the autocorrelation function of the process $U$, resulting from the summation of $N$ signals $u_i$ (not necessarily time dependent) is described by

$$\psi(\theta)_U = \lim_{T \to \infty} \frac{1}{T} \sum_{i=1}^{N} \sum_{j=1}^{N} \int_{-T/2}^{T/2} u_i(t)\, u_j(t-\theta)\, dt, \qquad (1)$$

where the terms with $i = j$ denote the autocorrelation functions of the originating signals, while the terms with $i \neq j$ represent the cross-correlations of these signals.

If the signals $u_i$ are similar, stationary and stochastic they furnish identical autocorrelation functions $\psi_a(\theta)$ and identical cross-correlation functions $\psi_c(\theta)$. Hence, the autocorrelation function of the resulting process $U$, will be $\psi(\theta)_U = N\psi_a(\theta) + N(N-1)\psi_c(\theta)$. Averagely, due the randomness of the signals, $\psi_c(\theta) = \langle u_i u_j \rangle = \langle u_i \rangle^2$.

The Wiener-Khintchine theorem permits us to find the power spectrum $S(\omega)_U$ of $U$. So, we have $S(\omega)_U = N \Im[\psi_a(\theta)] + N(N-1) \Im[\psi_c(\theta)]$, where the operator $\Im$ represents the Fourier transform and $\omega$ the angular frequency.

Therefore, we can write $S(\omega)_U = NS(\omega) + N(N-1)\langle u_i \rangle^2 \delta(\omega)$, where $S(\omega)$ is the power spectrum of the originating signals and $\delta(\omega)$ is the Dirac-delta.

The result $\psi_c(\theta) = \langle u_i \rangle^2$ cannot be accepted in particular cases. Indeed, when similar originating single-sided RTSs are summed, the cross-correlation terms, being due to the presence of the coincidences among the RTSs, are $\theta$-depending. For instance, if the coincidences in up level are missing for a particular $\theta$, or shift between the RTSs, $\psi_c(\theta)$ becomes zero, being zero the down-level of the RTSs.

The interevent time $\tau_k$ between two successive rising (or falling) edges of an RTS fluctuates around an average time $\langle \tau_k \rangle = \tau_u + \tau_d$, where $\tau_u$ denotes the mean lifetime in up level and $\tau_d$ that in down level. Therefore $\tau_k = t_k - t_{k-1}$, being $t_k$ the occurrence time of k-th rising (or falling) edge of the RTS. As a consequence, the time $t_k$ undergoes one-dimension Brownian increments.

About the relaxation time $\tau$, it is valid the relationship $1/\tau = 1/\tau_u + 1/\tau_d$. Hence, RTSs with $\tau_u = \tau_d$ have necessarily the relaxation time $\tau = \tau_u/2$, while if $\tau_u \ll \tau_d$, $\tau \cong \tau_u$ and vice versa.

The Lorentzian power spectrum of a RTS is of the form [1]:

$$S(\omega) = (\text{var } u)\frac{4\tau}{1+\omega^2\tau^2}, \qquad (2)$$

where $(\text{var } u)$ represents the variance of the signal. The asymptotic trend of Eq. (2) is as $1/\omega^2$ at high frequencies and constant at low frequencies, being $\omega = 1/\tau$ the corner frequency.

Fig. 1 shows a sketch of resulting signal due to the sum of few similar single-sided RTSs, where the self-organized shape of the pulses of coincidence, and the self-similarity typical of time series with $1/f$ power spectrum, are emerging.

It is also evident, in Fig. 1, that a fluctuation can dominate the neighbors of minor amplitude, so a dominant fluctuation tends to repeat itself at time intervals of mean $<\tau_k>$ in a train of discrete duration $n<\tau_k>$ (with $n$ integer). Indeed, due the Brownian spread of the occurrence times $t_k$ of the originating RTSs, the dominance of such fluctuations degrades in time, so the duration of the associate trains depends on the amplitude of dominant fluctuation and also on the competition with neighboring dominant fluctuations. It is important point out that the time interval between the pulses of such trains is not exactly $<\tau_k>$, but it is self-adjusting to $<\tau_k>$. In other words, the sequence of these pseudo-periods is not subjected to a Brownian diffusion.

About the power spectrum of resulting signal, the shape of pulses mainly influence the high frequencies of the spectrum, while the spectrum at low frequencies depends only by the interevent times [6][7].

Fig. 2 shows an example of resulting signal restricted only to the analysis of the correlations between the interevent times of the pulse-trains of various duration, without consider shape and amplitude of the pulses themselves.

The autocorrelation function of the two-level resulting signal of Fig. 2a has various relaxation times, corresponding to the durations of the pulse-trains, while the interruptions between the pulse-trains can be considered of white noise, which is uncorrelated. Therefore, we can decompose the resulting signal in a set of independent Lorentzians of various relaxation times (Fig.

2b), being the independence of these Lorentzians due to the lack of coincidences, and then of cross-correlations.

Looking to the set of Lorentzians of Fig. 2b, we note that increasing $\tau$ increases of the rarity of the fluctuations. Hence, imaging a continuous distribution of relaxation times [1], putting $(\text{var} u) \propto 1/\tau$ in the interval $\tau_2^{-1} \ll \tau^{-1} \ll \tau_1^{-1}$ and zero outside this interval, one has $S(\omega) \propto 1/\omega$ in the frequency range $\tau_2^{-1} \ll \omega \ll \tau_1^{-1}$, while the spectral components, due to the pseudo-periodicity of the pulses, pertain the high frequencies, outside the range interested by $1/f$ noise.

In conclusion, the model of $1/f$ noise, ascribed to a superposition of Lorentzians with a distribution of relaxation times, appears not only as a consequence of a sum of similar single-sided RTSs with the same relaxation time, but becomes useful to demonstrate the appearance of $1/f$ noise. In such a manner the discussed problem of the necessity of a distribution of relaxation times, and of the independence of the Lorentzians [4], finds its solution. To further support of the appearance of $1/f$ noise, this has been analytically demonstrated in a resulting signal formed by an unique sequence of pseudo-periodic pulses with a period self-adjusting to $<\tau_k>$ (the interruptions between the trains are missing) [6].

Besides, the presence of a Lorentzian together with $1/f$ noise has been ascertained experimentally [8][9].

About a computer simulation, we have considered single-sided RTSs with $\tau_u = \tau_d$, selecting random, with a steady distribution, the transition between the two levels within the interval $\tau_k$ fluctuating with a Gaussian distribution. In practice we put $\tau_k = <\tau_k> + \sigma \varepsilon_k$, where $\varepsilon_k$ is a random variable normally distributed around the zero with unity variance and standard deviation $\sigma$.

The simulations was performed using MATLAB. Obviously, duration of the RTSs, and sampling time, limit the range of the power spectrum. This limit, and the number of summed RTS, involve the computation time which, in our case, has reached some hours.

Fig. 3 displays an expanded time series of a superposition of many single-sided RTSs, with same relaxation time, simulated by computer selecting $\tau_u = \tau_d = 2.5$, $\sigma = 1$, sampling time = 0.1, then $\tau = 1.25$ and $<\tau_k> = 5$, in arbitrary units. A train of pulses appears in Fig.3, with average period $<\tau_k> = 5$, in competition with other trains of pulses separated by intervals of uncorrelated noise.

Fig. 4 shows the power spectrum of a superposition of 50 single-sided similar RTSs, with the same relaxation time ($\tau_u = \tau_d = 2.5$, $\sigma = 1$, sampling time = 0.1, $\tau = 1.25$, $<\tau_k> = 5$, in arbitrary units, and duration = 800 interevent times $\tau_k$). This spectrum results $1/f$ at low frequencies and $1/f^2$ at high frequencies, as it is foreseen.

**Acknowledgements**

The author is indebted with Dr. Bruno Burrometo for simulations in MATLAB.

**Figure captions**

FIG. 1 . Graphic simulation of self-organized and self-similar waveforms in the sum of similar RTSs. Fig1a shows a time series representing the result of the superposition of the five RTS, with same relaxation time, which appear in Fig.1b. The resulting fluctuations are single-sided, due the low number of originating signals which does not permit the generation of symmetrical fluctuations around a mean level.

FIG. 2 . Graphic simulation of decomposition of a two-level process. The signal of Fig.2a, consisting of pulse-trains of various length, is decomposed in the set of independent RTSs of Fig.2b, having different relaxation times.

FIG. 3 . Pseudo-periodicity and pulse-trains in a time series derived from a sum of many similar single-sided RTSs. The figure shows the result of a computer simulation obtained summing many single-sided RTSs with the same relaxation time (the value of the used parameters is in the text). It is evident the competition between trains of pseudo-periodic pulses, as it appears from the interruption of the pseudo-periodicity.

FIG. 4 . Power spectrum of a sum of 50 similar single-sided RTSs with same relaxation time and simulated by computer (the value of the used parameters appears the text). It is evident the appearance of $1/f$ noise at low frequencies while the Lorentzian behavior remains at high frequencies. Truncation at low frequencies depends on the duration of the RTSs, while the limit at high frequencies depends on the sampling time

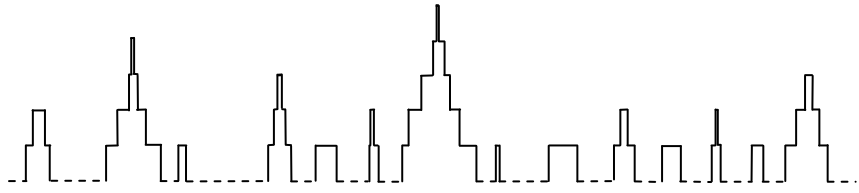

a

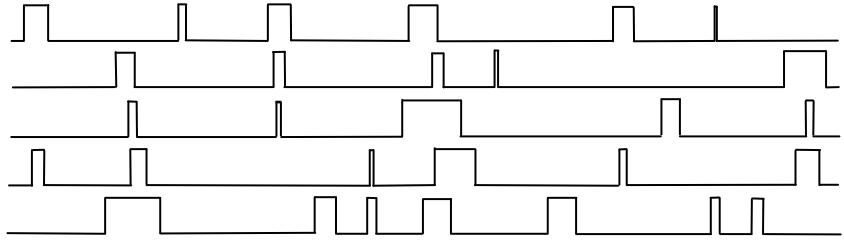

b

**FIG. 1**

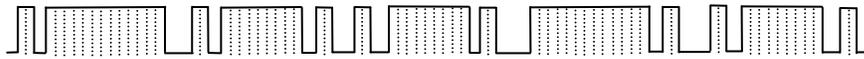

a

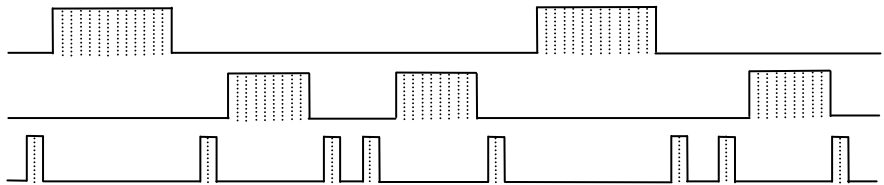

b

**FIG. 2**

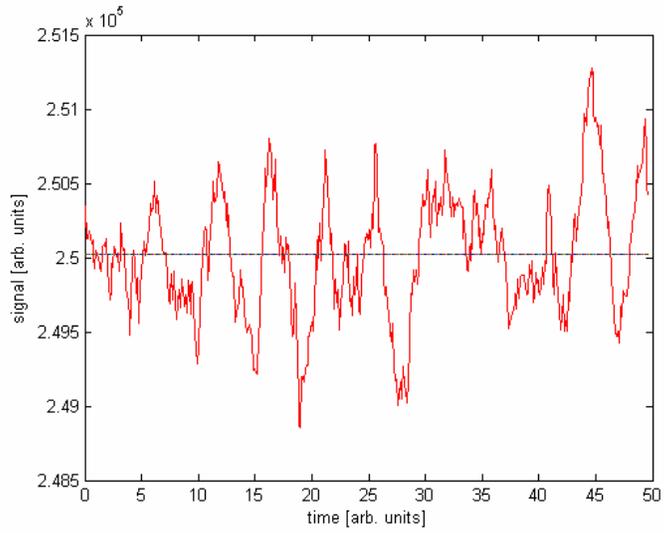

**FIG. 3**

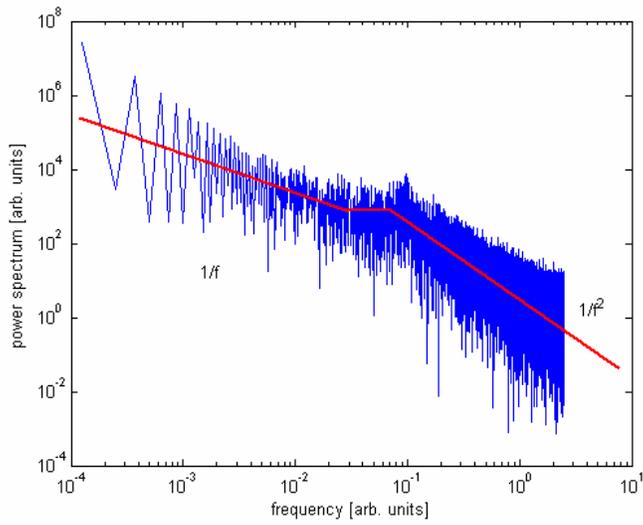

**FIG. 4**